\documentclass[aps,prl,twocolumn,amsmath,amssymb,groupedaddress,longbibliography
]{revtex4-2}

\usepackage{graphicx}
\usepackage{dcolumn}
\usepackage{bm}
\usepackage{multirow}
\usepackage{braket}
\usepackage{longtable}
\usepackage{color}
\usepackage{ulem}
\usepackage{booktabs}

\begin{document}

\title{
Time-reversal switching responses in antiferromagnets 
}

\author{Satoru Hayami$^1$ and Hiroaki Kusunose$^{2,3}$}
\affiliation{
$^1$Graduate School of Science, Hokkaido University, Sapporo 060-0810, Japan \\
$^2$Department of Physics, Meiji University, Kawasaki 214-8571, Japan \\
$^3$Quantum Research Center for Chirality, Institute for Molecular Science, Okazaki 444-8585, Japan
}

\begin{abstract}
We propose emergent time-reversal switching responses in antiferromagnets, which is triggered by an accompanying magnetic toroidal monopole, i.e., time-reversal odd scalar distinct from electric and magnetic monopoles.
We show that simple collinear antiferromagnets exhibit unconventional responses to external electric and/or magnetic fields once magnetic symmetry accommodates the magnetic toroidal monopole. 
We specifically demonstrate that the emergence of the magnetic toroidal monopole in antiferromagnets enables us to control rotational distortion by an external magnetic field, switch vortex-type antiferromagnetic structure by an external electric field, and convert right/left-handedness in chirality by a composite electromagnetic field. 
We also present the symmetry conditions to induce the magnetic toroidal monopole and exhibit candidate materials including noncollinear antiferromagnets in order to stimulate experimental observations. 
\end{abstract}

\maketitle

{\it Introduction.---}
Monopole is the most fundamental object in electromagnetism. 
An electric (magnetic) monopole $Q_0$ ($M_0$) corresponds to an elementary electric (magnetic) charge. 
Although the magnetic monopole as an elementary particle has never been observed so far, extended objects with the same symmetry have been found in condensed matter physics in the context of spin ice~\cite{castelnovo2008magnetic, morris2009dirac, bramwell2009measurement},  multiferroics~\cite{Spaldin_PhysRevB.88.094429, khomskii2014magnetic}, and topological insulators~\cite{qi2009inducing, Martin_PhysRevD.98.056012, Meier_PhysRevX.9.011011}.

The electric (magnetic) monopole is characterized by a time-reversal ($\mathcal{T}$) even scalar ($\mathcal{T}$-odd pseudoscalar) with respect to the space-time inversion. 
One can also introduce their counterparts with opposite parities: 
an electric toroidal monopole $G_0$ corresponding to the $\mathcal{T}$-even pseudoscalar and a magnetic toroidal monopole (MTM) $T_0$ corresponding to the $\mathcal{T}$-odd scalar~\cite{kusunose2022generalization}. 
Their practical representation can be made based on the symmetry-adapted multipole basis that constitutes a complete basis set~\cite{Kusunose_PhysRevB.107.195118}.
Recently, the former $G_0$ has been recognized as a microscopic physical quantity to characterize the chirality~\cite{Hayami_PhysRevB.98.165110,kishine2022definition}, which becomes the origin of the cross-correlation phenomena between polar and axial quantities, such as current-induced magnetization (Edelstein effect)~\cite{edelstein1990spin} and electric-field-induced rotational distortion~\cite{Oiwa_PhysRevLett.129.116401}. 
On the other hand, the latter MTM has been still an enigmatic monopole, whose realization and physical nature have been unclear. 

\begin{figure}[t!]
\begin{center}
\includegraphics[width=0.9 \hsize ]{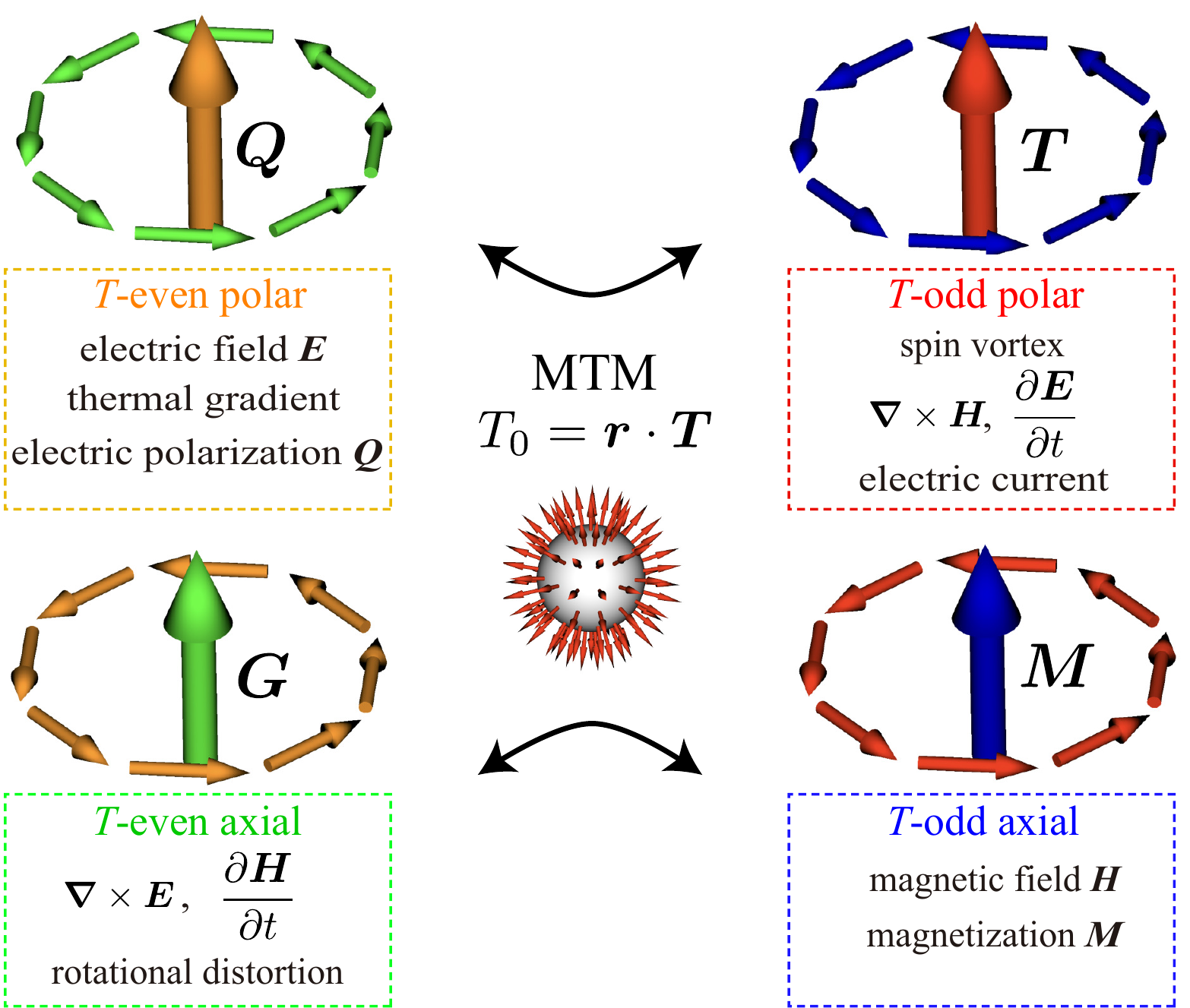} 
\caption{
\label{fig: ponti}
Conversions among different dipoles in terms of magnetic toroidal monopole $T_0$ defined by an inner product of the position vector $\bm{r}$ and magnetic toroidal dipole $\bm{T}$. 
The electric dipole $\bm{Q}$ (electric toroidal dipole $\bm{G}$) denoted by the orange (green) arrow can be converted to $\bm{T}$ magnetic dipole $\bm{M}$) denoted by the red (blue) arrows via $T_0$, and vice versa. 
Related representative vectors are shown in each lower panel.
}
\end{center}
\end{figure}

In the present study, we theoretically propose the emergent MTM in antiferromagnets and elucidate electromagnetic responses driven by its ordering. 
We show that the MTM gives rise to a variety of time-reversal switching responses between polar (axial) quantities, such as magnetic-field-induced rotational distortion, electric-field-induced spin vortex, and electromagnetic-field-induced chirality. 
Moreover, we show all the magnetic point groups to accommodate the MTM and exhibit candidate materials in both collinear and noncollinear antiferromagnets. 
We also demonstrate such physical phenomena under the MTM ordering by considering a minimal collinear antiferromagnetic model, and propose a possible optical rotation measurement in Ca$_{2}$RuO$_{4}$. 
Our results provide a guideline to search for unconventional antiferromagnets with the MTM.

{\it Magnetic toroidal monopole.---}
Magnetic toroidal multipole is characterized by a $\mathcal{T}$-odd polar tensor, which shows a different spatial parity from magnetic multipole~\cite{dubovik1975multipole, dubovik1990toroid, Spaldin_0953-8984-20-43-434203, kopaev2009toroidal, hayami2018microscopic}. 
Among them, the dipole component, i.e., magnetic toroidal dipole $\bm{T}$, which is expressed as a vector product of the magnetic dipole $\bm{M}$ (or spin $\bm{S}$) and the position vector, i.e., $\bm{T} \propto \bm{r} \times \bm{M} (\bm{S})$ [upper-right panel of Fig.~\ref{fig: ponti}], has been extensively studied, since it leads to the linear magnetoelectric effect~\cite{schmid2001ferrotoroidics, van2007observation, Spaldin_0953-8984-20-43-434203, kopaev2009toroidal, Hayami_PhysRevB.90.024432, hayami2016emergent, thole2018magnetoelectric} and nonreciprocal transport~\cite{Kubota_PhysRevLett.92.137401, Sawada_PhysRevLett.95.237402, tokura2018nonreciprocal, Gao_PhysRevB.97.134423, Watanabe_PhysRevResearch.2.043081, Yatsushiro_PhysRevB.105.155157, Wang_PhysRevLett.127.277201, Kirikoshi_PhysRevB.107.155109}.  
By using $\bm{T}$, the MTM ($T_0$) is expressed as 
\begin{align}
\label{eq: MTM}
T_0 = \bm{r} \cdot \bm{T}. 
\end{align}
The schematic picture of $T_0$ is shown in middle of Fig.~\ref{fig: ponti}. 
Although $T_0$ identically vanishes in the single atomic wave function owing to $\bm{r} \perp \bm{T}$~\cite{hayami2018microscopic, Hayami_PhysRevB.98.165110}, it survives in a magnetic cluster like antiferromagnets, as discussed below. 
Note that $T_0$ is totally independent of the other three monopoles ($Q_0$, $M_0$, and $G_0$), which have orthogonal matrix elements to $T_{0}$.

{\it Cross-correlation phenomena.---}
Since $T_0$ is a $\mathcal{T}$-parity opposite to an electric charge, it plays a role to convert between two polar-vector quantities with opposite $\mathcal{T}$ parity. 
Considering that $\bm{r}$ is symmetry-equivalent to the electric dipole $\bm{Q}$, one can find a correspondence between $T_0$, $\bm{Q}$, and $\bm{T}$ from Eq.~(\ref{eq: MTM}) as 
\begin{align}
T_0 \leftrightarrow \bm{Q} \cdot \bm{T}, 
\end{align}
where $\bm{Q}$ and $\bm{T}$ corresponds to $\mathcal{T}$-even and $\mathcal{T}$-odd polar vectors, respectively. 
Similarly, noting the relation of $\bm{T} \propto (\bm{Q}\times \bm{M})$, Eq.~(\ref{eq: MTM}) is rewritten as 
\begin{align}
T_0 \leftrightarrow \bm{G} \cdot \bm{M}, 
 \end{align}
where $\bm{G}=\bm{r}\times \bm{Q}$ represents an electric toroidal dipole corresponding to a $\mathcal{T}$-even axial vector~\cite{dubovik1986axial, Hlinka_PhysRevLett.113.165502, jin2020observation, hayashida2020visualization, Hayami_doi:10.7566/JPSJ.91.113702, inda2023nonlinear}. 
Thus, $T_0$ can also convert between two axial-vector quantities with opposite $\mathcal{T}$ parity. 
The conversion properties among dipoles ($\bm{Q}, \bm{M}, \bm{T}, \bm{G}$) via $T_{0}$ are summarized in Fig.~\ref{fig: ponti}; we also show representative vector quantities in each lower panel. 

The above symmetry argument indicates emergent time-reversal switching responses under the MTM ordering, e.g., the free energy is expanded by the electric field $\bm{E}$ and magnetic field $\bm{H}$ in addition to the conventional term $F_{0}$ as 
\begin{align}
\label{eq: free}
F = &F_0 - \alpha_1 \bm{G} \cdot  \bm{H} - \alpha_2  \bm{T} \cdot \bm{E}\cr 
&- \alpha_3 \bm{Q} \cdot (\bm{\nabla}\times \bm{H}) - \alpha_4 \bm{M} \cdot (\bm{\nabla}\times \bm{E}) +  \cdots, 
\end{align}
where $\alpha_1$--$\alpha_4$ are coefficients, which can be finite only when the thermal average of $T_{0}$ is finite. 
It is noted that $\mathcal{T}$-opposite $\bm{H}$, $\bm{E}$, $\bm{\nabla}\times \bm{H}$, and $\bm{\nabla}\times \bm{E}$ become the conjugate fields of $\bm{G}$, $\bm{T}$, $\bm{Q}$, and $\bm{M}$, respectively. 
Especially, $\bm{\nabla}\times \bm{H}$ and $\bm{\nabla}\times \bm{E}$ correspond to the rotational distortion in terms of the spin and charge degrees of freedom, respectively, and have the same symmetry as the electric current and time derivative of $\bm{H}$.
Thus, unusual cross-correlation responses occur in the presence of $T_{0}$ under external fields; a homogeneous magnetic (electric) field gives rise to $\bm{G}$ ($\bm{T}$) corresponding to the vortex of $\bm{Q}$ ($\bm{M}$), while an inhomogeneous magnetic (electric) field or electric current (time derivative of magnetic field) with finite rotation leads to the electric polarization (magnetization).
Accordingly, one can experimentally control the rotational distortion by applying $\bm{H}$, switch the vortex-type antiferromagnetic domain by $\bm{E}$, and the favorite handedness of induced chirality $G_{0}$ by $\bm{E}\cdot\bm{H}$, as demonstrated below.

\begin{table}[t!]
\caption{
Classification of point groups accompanying order parameters of $(T_0, M_z, T_z, M_0)$.
The candidate materials are also listed.  
The subscripts $m,n$ in the point group stand for $n=2,3,4,6$ and $m=2,3$. 
}
\label{tab: mp}
\centering
\renewcommand{\arraystretch}{1.0}
 \begin{tabular}{llllccc}
 \hline  \hline
Point group &  $T_0$ & $M_z$  & $T_z$  & $M_0$  & Materials     \\ \hline 
$O_{\rm h}$, $T_{\rm d, h}$, $D_{n{\rm h}, m{\rm d}}$ & $\checkmark$ & -- & -- & -- & KMnF$_3$\cite{knight2020nuclear}, Ca$_2$RuO$_4$\cite{Porter_PhysRevB.98.125142} \\
$C_{n{\rm h}}$, $S_4$, $C_{3{\rm i}}$, $C_{\rm i}$ & $\checkmark$ & $\checkmark$ & -- & -- & MnV$_2$O$_4$\cite{Garlea_PhysRevLett.100.066404}, Mn$_3$As$_2$\cite{karigerasi2021two} \\
$C_{n{\rm v}}$ & $\checkmark$ & -- & $\checkmark$ & --  & YMnO$_3$\cite{Munoz_PhysRevB.62.9498}, Er$_2$Cu$_2$O$_5$\cite{Garca-Muoz_PhysRevB.44.4716}\\
$O$, $T$, $D_n$ & $\checkmark$ & -- & -- & $\checkmark$ &  Ho$_2$Ge$_2$O$_7$\cite{Morosan_PhysRevB.77.224423}, Mn$_3$IrGe\cite{eriksson2004structural}\\
$C_{\rm s}$  & $\checkmark$ & $\checkmark$ & $\checkmark$ & -- & Mn$_4$Nb$_2$O$_9$\cite{solana2021mn} \\
$C_{n}$, $C_1$ & $\checkmark$ & $\checkmark$ & $\checkmark$ & $\checkmark$ & ScMnO$_3$\cite{Munoz_PhysRevB.62.9498}, Mn$_2$FeMoO$_6$\cite{li2014magnetic}\\
 \hline
\hline 
\end{tabular}
\end{table}

{\it Symmetry conditions.---}
Let us discuss the symmetry condition to accormodate the MTM. 
Since the MTM is equivalent to a $\mathcal{T}$-odd scalar without spatial anisotropy, the necessary symmetry breaking is only the $\mathcal{T}$ symmetry with keeping the original point group symmetry~\cite{Yatsushiro_PhysRevB.104.054412}. 
Among 122 magnetic point groups, 32 crystallographic point groups without $\mathcal{T}$ operation satisfy this condition, as summarized in Table~\ref{tab: mp}~\cite{bradley2009mathematical}. 

Moreover, we classify the above 32 point groups into 6 types accompanying the activation of the $z$-component magnetic dipole $M_z$, $z$-component of the magnetic toroidal dipole $T_z$, and magnetic monopole $M_0$, as shown in Table~\ref{tab: mp}. 
When considering the point groups where $M_z$ belongs to the totally symmetric irreducible representation, i.e., $C_{n{\rm h}}$, $S_4$, $C_{3{\rm i}}$, $C_{\rm i}$, $C_{\rm s}$, $C_{n}$, and $C_1$ ($n=2,3,4,6$), one can control the MTM domain by using $H_z$. 
In the case of $C_{n{\rm v}}$, $C_{\rm s}$, $C_{n}$, and $C_1$ with $T_z$, applying the electric field enables us to select the MTM domain. 
For $O$, $T$, $D_n$, $C_{n}$, and $C_1$ with $M_0$, a further cross-correlation response between polar and axial quantities, e.g., $\bm{Q}\,\leftrightarrow\,\bm{G}$ and $\bm{Q}\,\leftrightarrow\, \bm{M}$, is expected like enantiomorphic point groups. 
Lastly, the point groups, $O_{\rm h}$, $T_{\rm d}$, $T_{\rm h}$, $D_{n{\rm h}}$, and $D_{m{\rm d}}$  $(m=2,3)$, accompany neither $M_z$, $T_z$, nor $M_0$, whose system exhibits a pure MTM and its related physical responses. 

The MTM can be realized by antiferromagnetic phase transitions satisfying the above symmetry condition. 
We exhibit candidate antiferromagnetic materials accompanying the MTM in Table~\ref{tab: mp}, which are referred from MAGNDATA~\cite{gallego2016magndata}, magnetic structures database. 
Various materials possess the MTM irrespective of the lattice and antiferromagnetic structures, e.g., collinear magnetic structure under the tetragonal point group KMnF$_3$~\cite{knight2020nuclear} and noncollinear magnetic structure under the cubic point group Mn$_3$IrGe~\cite{eriksson2004structural}. 
In these materials, physical phenomena characteristic of the MTM, such as the magnetic-field-induced rotational distortion and electric-field-induced spin vortex, can be expected. 
We show several antiferromagnetic structures to accommodate the MTM under different point groups in Supplemental Material~\cite{SM_MTM}.

\begin{figure}[t!]
\begin{center}
\includegraphics[width=1.0 \hsize ]{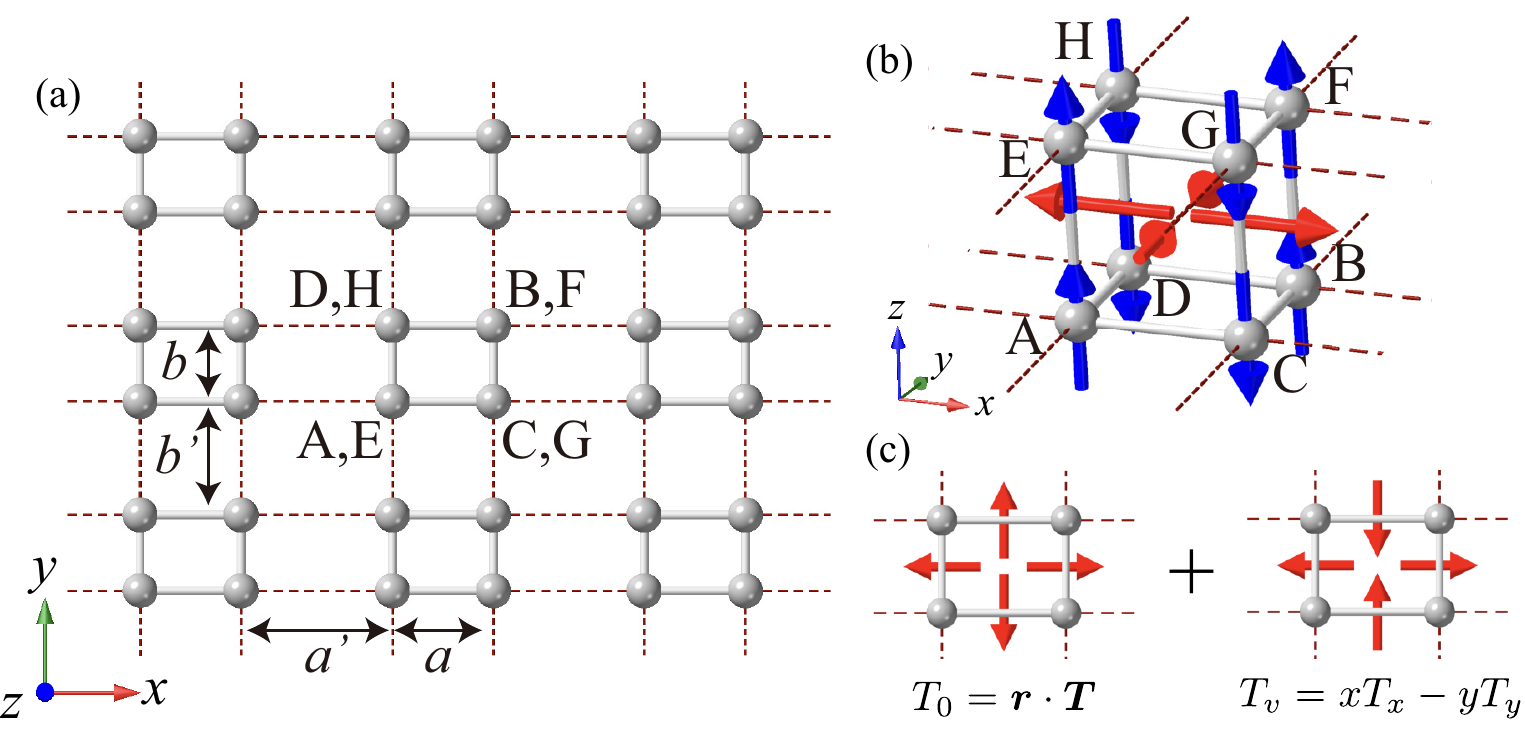} 
\caption{
\label{fig: lattice}
(a) Orthorhombic crystal structure with sublattices A--H. 
(b) Collinear magnetic ordering accompanying $T_0$, where the blue and red arrows represent the spin and $\bm{T}$, respectively. 
In (b), the outward and inward red arrows have different lengths. 
(c) The distribution of $\bm{T}$ in (b) is decomposed into $T_0$ and quadrupole component $T_v$. 
}
\end{center}
\end{figure}

{\it Model calculations.---}
To demonstrate the role of the MTM in antiferromagnets and its cross-correlation coupling in Eq.~(\ref{eq: free}), we analyze a minimal $s$-$p$ model; the physical space spanned by four orbitals and spin includes all the dipoles ($\bm{Q}, \bm{G}, \bm{M}, \bm{T}$), which needs to describe physical responses in Eq.~(\ref{eq: free})~\cite{kusunose2020complete, Hayami_PhysRevB.106.144402}. 
It is noted that the following results are not qualitatively altered by choosing different orbitals and lattice structures, once the relevant multipole degrees of freedom, such as ($\bm{Q}, \bm{G}, \bm{M}, \bm{T}$), are included in the low-energy physical space.
We consider a bilayer lattice structure consisting of a cuboid with eight sublattices A--H under the space group $Pmmm$ ($D^1_{\rm 2h}$), as shown in Fig.~\ref{fig: lattice}(a); we set the unit of lattice constants as $a=a'=b=b'=0.5$ and $c=1$ ($c$ is the bond length between sublattices A and E) without loss of generality. 
The Hamiltonian is given by 
\begin{multline}
\label{eq: Ham}
\mathcal{H}=\sum_{\substack{\bm{k} \gamma\alpha\sigma \\ \gamma'\alpha'\sigma'}}c^{\dagger}_{\bm{k}\gamma\alpha\sigma} 
( \delta_{\sigma\sigma'}H^{t}
+
 \delta_{\gamma\gamma'} 
H^{{\rm SOC}}
+
\delta_{\alpha\alpha'}H^{{\rm M}})
c_{\bm{k}\gamma'\alpha'\sigma'},  
\end{multline}
where $c^{(\dagger)}_{\bm{k}\gamma\alpha\sigma}$ represents the annihilation (creation) operator of electrons at wave vector $\bm{k}$, sublattice $\gamma=$ A--H, orbital $\alpha=s$, $p_x$, $p_y$, and $p_z$, and spin $\sigma$. 
In Eq.~(\ref{eq: Ham}) $H^{t}$ includes the nearest-neighbor hopping for the intra- and inter-unit cuboid. 
We adopt the Slater-Koster parameter for the intra-cuboid hoppings: for the $x$-bond direction, $t^x$ for the hopping between $s$ orbitals ($\alpha,\alpha'=s$), $t^x_p$ for that between $(p_x, p_y)$ orbitals ($\alpha,\alpha'=p_x, p_y$), $t^x_z$ for that between $p_z$ orbitals ($\alpha,\alpha'=p_z$), and $t^x_{sp}$ for that between different $s$-$(p_x, p_y)$ orbitals ($\alpha=s$ and $\alpha'=p_x, p_y$ and vice versa). 
We regard $t^x=-1$ as the energy unit of the model and set $t^x_{p}=0.7$, $t^x_z=0.2$, and $t^x_{sp}=0.3$. 
Similarly, we set the intra-cuboid hoppings along the $y$ and $z$ directions by multiplying 0.9 and 0.5 by that along the $x$ direction. 
In addition, we set the inter-cuboid hoppings along the $x$ and $y$ directions by multiplying 0.8 by intra-cuboid ones. 
It is noted that the choice of the hopping parameters does not affect the following results qualitatively. 
$H^{\rm SOC}$ in Eq.~(\ref{eq: Ham}) means the atomic spin--orbit coupling for three $p$ orbitals with $\lambda=0.5$. 

$H^{\rm M}$ in the third term in Eq.~(\ref{eq: Ham}) denotes the mean-field term to describe the antiferromagnetic ordering. 
We consider the collinear antiferromagnetic ordering in Fig.~\ref{fig: lattice}(b), where $H^{\rm M}$ is explicitly given by 
\begin{align}
H^{\rm M}=&
-h \delta_{\gamma \gamma'} p(\gamma) \sigma_z. 
\end{align}
Here, $p(\gamma)=+1 (-1)$ for sublattices A, B, E, and F (C, D, G, and H), and $\sigma_z$ represents the $z$-component Pauli matrix in spin space. 
We set the amplitude of antiferromagnetic molecular field as $h=2$ and consider the low-electron filling per site $n_{\rm e}=(1/N)\langle \sum_{\bm{k}\gamma\alpha\sigma}c^{\dagger}_{\bm{k}\gamma\alpha\sigma}c^{}_{\bm{k}\gamma\alpha\sigma}\rangle=0.2$, where $N=8\times1600^2$ is the total sites and $n_{\rm e}=8$ (four orbitals times two spins) represents the full filling.

The eight-sublattice collinear magnetic structure in Fig.~\ref{fig: lattice}(b) satisfies the symmetry condition to accommodate the MTM; inversion, three two-fold rotation, and three mirror symmetries under the space group $Pmmm$ remain and only the time-reversal symmetry is broken. 
Indeed, by closely looking into the collinear spin configuration denoted by the blue arrows in Fig.~\ref{fig: lattice}(b) on each plaquette of the cuboid, $\bm{T}$, which is defined by the vector product of spins and the position vector measured from the center of each plaquette, becomes nonzero for the sides: the outward $x$ component of $\bm{T}$ emerges on the plaquettes ADHE and CBFG and the inward $y$ component emerges on the plaquettes ACGE and DBFH, as shown by the red arrows in Fig.~\ref{fig: lattice}(b). 
Since the $xz$ and $yz$ planes are inequivalent in the orthorhombic structure, the amplitudes of the $x$ and $y$ components of $\bm{T}$ are different from each other. 
The distribution of $\bm{T}$ in Fig.~\ref{fig: lattice}(b) is decomposed into the linear combination of $T_0$ and quadrupole component $T_v=x T_x - y T_y$ as shown in Fig.~\ref{fig: lattice}(c), which means a net component of $T_0$ in the unit cuboid. 
In this way, the collinear antiferromagnetic structure in Fig.~\ref{fig: lattice}(b) accompanies the MTM. 
Similar collinear magnetic structures have been identified in materials, such as Fe$_2$PO$_5$~\cite{warner1992valence}, $X$CrO$_3$ ($X=$ Sc, In, Tl, La)~\cite{martinelli2011crystal, Ding_PhysRevB.95.054432}, and $Y$FeO$_3$ ($Y=$ Ce, Nd, Dy)~\cite{plaza1997neutron, ritter2021determination, ritter2022magnetic}; these materials are the potential candidates hosting the MTM.

\begin{figure}[t!]
\begin{center}
\includegraphics[width=1.0 \hsize ]{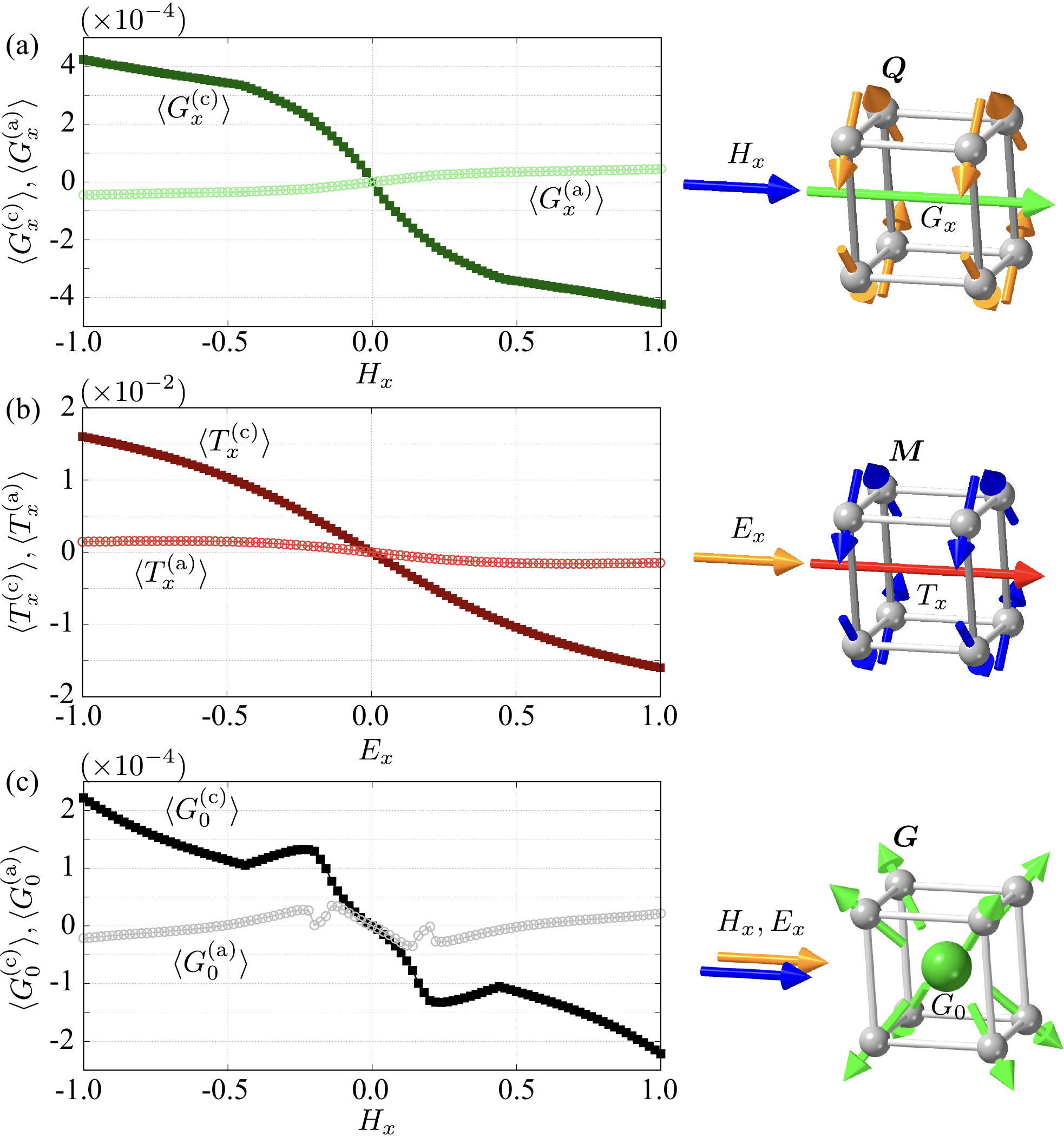} 
\caption{
\label{fig: data}
(a) Magnetic field $H_x$ dependence of the electric toroidal dipole $\langle G_x \rangle $. 
(b) Electric field $E_x$ dependence of the magnetic toroidal dipole $\langle T_x \rangle $. 
(c) $H_x$ dependence of the electric toroidal monopole $\langle G_0 \rangle $ in the presence of $E_x=0.2$. 
The right panel in each figure shows the schematic pictures corresponding to the left panel. 
The green sphere represents $\langle G_0 \rangle $. 
}
\end{center}
\end{figure}

Using the model in Eq.~(\ref{eq: Ham}), we demonstrate the cross-correlation phenomena originating from the effective coupling in Eq.~(\ref{eq: free}).
First, we discuss the magnetic-field-induced rotational distortion by introducing the Zeeman Hamiltonian $\mathcal{H}^{\rm Z}$ coupled to spin as $\mathcal{H}^{\rm Z}= -H_x\sum_{\bm{k}\gamma \alpha\sigma\sigma'}c^{\dagger}_{\bm{k}\gamma \alpha \sigma} \sigma^x_{\sigma\sigma'} c^{}_{\bm{k}\gamma \alpha \sigma'}$. 
Since the microscopic degree of freedom corresponding to the rotational distortion is $\bm{G}$, we calculate its expectation values in the atomic and cluster forms, $\langle G^{\rm (a)}_x \rangle$ and $\langle G^{\rm (c)}_x \rangle$, against the applied magnetic field $H_x$~\cite{Kusunose_PhysRevB.107.195118}. 
Here, $G^{\rm (a)}_x$ is the atomic-scale definition using $(\bm{l} \times \bm{\sigma})_x$ ($\bm{l}$ is the orbital angular momentum) and $G^{\rm (c)}_x$ is the cluster definition formed by the vortex of the local electric dipoles by the orange arrows in Fig.~\ref{fig: data}(a); see Supplemental Material~\cite{SM_MTM} for the detailed expressions. 
As shown in the left panel of Fig.~\ref{fig: data}(a), both quantities become nonzero for $H_x \neq 0$; their sign is reversed by reversing the magnetic-field direction. 
This response coming from the interband process is non-dissipative within the linear response, which occurs in both metals and insulators. 
We also discuss the order parameter dependence and the magnetic-field-induced rotational distortion for noncollinear spin textures in Supplemental Material~\cite{SM_MTM}. 

Next, let us consider the electric-field-induced spin vortex (the time-reversal counterpart of the previous example), where $\bm{T}$ is induced along the external electric-field direction. 
We introduce the local $s$-$p_x$ hybridized Hamiltonian as $\mathcal{H}^{\rm E}= -E_x\sum_{\bm{k}\gamma\sigma}(c^{\dagger}_{\bm{k}\gamma s   \sigma} c^{}_{\bm{k}\gamma p_x  \sigma}+{\rm H.c.})$ corresponding to the coupling between the electric dipole moment and the applied electric field. 
Figure~\ref{fig: data}(b) shows the $E_x$ dependence of the atomic contribution of $\bm{T}$, $\langle T^{\rm (a)}_x \rangle$, and the cluster one, $\langle T^{\rm (c)}_x \rangle$; $T^{\rm (a)}_x$ is represented by the local imaginary $s$-$p_x$ hybridization and $T^{\rm (c)}_x$ is represented by the spin vortex, as shown in the right panel of Fig.~\ref{fig: data}(b)~\cite{SM_MTM}. 
Similarly to Fig.~\ref{fig: data}(a), both $\langle T^{\rm (a)}_x \rangle$ and $\langle T^{\rm (c)}_x \rangle$ become nonzero for $E_x \neq 0$, and their sign is reversed when the sign of $E_x$ is changed. 
Thus, the spin vortex can be switched by applying the electric field under the MTM ordering. 
This response also arises from the non-dissipative interband process within the linear response. 

Furthermore, we find that the system acquires the chirality, i.e., a finite expectation value of $G_{0}$, when both static $H_x$ and $E_x$ are applied simultaneously. 
We show the behaviors of atomic-scale and cluster electric toroidal monopoles, $\langle G^{\rm (a)}_0 \rangle$ and $\langle G^{\rm (c)}_0 \rangle$, in Fig.~\ref{fig: data}(c), which are the microscopic measure of chirality; the former is described by the atomic spin-dependent imaginary $s$-$p$ hybridization and the latter is described by the source of the $\bm{G}$ flux in the cuboid~\cite{SM_MTM}. 
As shown in Fig.~\ref{fig: data}(c), the result indicates $\langle G^{\rm (a)}_0 \rangle$ and $\langle G^{\rm (c)}_0 \rangle$ are induced by $H_x$, $E_x \neq 0$, and their sign is reversed when the direction of either $H_{x}$ or $E_{x}$ is reversed. 
This result is consistent with the symmetry of the system in the presence of $H_x$ and $E_x$; there are no inversion and mirror symmetries. 
It is noted that $\langle G^{\rm (a,c)}_0 \rangle=0$ in the paramagnetic $Pmmm$ system without $T_0$ under a nonconjugate field of $G_{0}$, $H_x$ and $E_x$, since the time-reversal parity of $H_x E_x$ is opposed to that of $G_0$. 
In other words, the induction of $\langle G_0 \rangle $ by the composite field $H_x E_x$ is one of the characteristic features of the MTM ordering.

\begin{figure}[t!]
\begin{center}
\includegraphics[width=1.0 \hsize ]{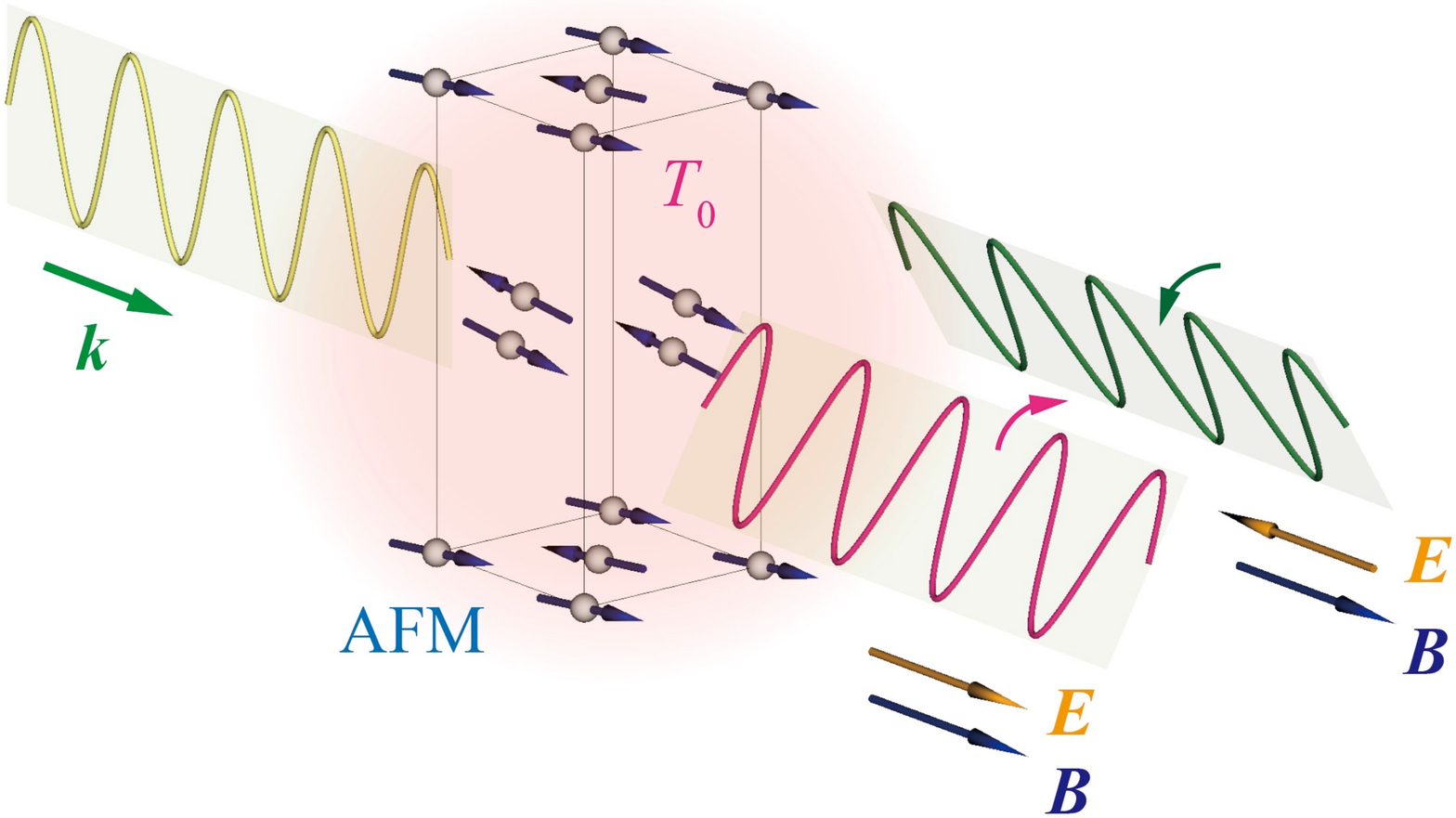} 
\caption{
\label{fig: optical_rotation}
Optical rotation in the antiferromagnet Ca$_2$RuO$_4$ with $T_0$ under the static electromagnetic field. 
$\bm{k}$ is the incident wave vector. 
The opposite rotations occur for the parallel and anti-parallel application of $\bm{E}$ and $\bm{H$}.
}
\end{center}
\end{figure}

{\it Conclusion.---}
We proposed the time-reversal odd scalar order parameter, i.e., the MTM ($T_{0}$), in antiferromagnets. 
We found that the MTM becomes a source of various time-reversal switching responses, such as magnetic-field-induced rotational distortion, electric-field-induced spin vortex, and electromagnetic-field-induced chirality, which are qualitatively different from other known multipole orderings like magnetic monopole and magnetic toroidal dipole. 
Furthermore, we showed the symmetry condition of the MTM as well as the candidate materials. 
Finally, we demonstrated the minimal model to host the MTM in collinear antiferromagnets.

In order to stimulate findings of cross-correlation physical phenomena driven by the MTM, we propose an experimental setup in a candidate noncollinear antiferromagnet Ca$_2$RuO$_4$, which accompanies a pure MTM in Table~\ref{tab: mp}~\cite{SM_MTM}, by focusing on the optical rotation inherent in chirality. 
Since the sign of $\langle G_0 \rangle$, i.e., handedness of chirality, is determined by the relative direction of electric and magnetic fields, as shown in Fig.~\ref{fig: data}(c), one can expect the switching of right-and left-handed rotations by reversing one of the fields, as schematically shown in Fig.~\ref{fig: optical_rotation}. 
In addition, the other cross-correlation phenomena proposed above, such as rotational distortion by an external magnetic field and induction of the vortex-type antiferromagnetic structure by an external electric field, are also expected.

\begin{acknowledgments}
This research was supported by JSPS KAKENHI Grants Numbers JP21H01031, JP21H01037, JP22H04468, JP22H00101, JP22H01183, JP23K03288, JP23H04869, JP23H00091 and by JST PRESTO (JPMJPR20L8).
Parts of the numerical calculations were performed in the supercomputing systems in ISSP, the University of Tokyo.
\end{acknowledgments}

\bibliographystyle{apsrev}
\bibliography{ref}

\end{document}